\documentclass[letterpaper]{JHEP}
\usepackage{epsfig}

\newcommand{\be}{\begin{equation}}
\newcommand{\ee}{\end{equation}}
\newcommand{\bea}{\begin{eqnarray}}
\newcommand{\eea}{\end{eqnarray}}

\newcommand{\IZ}{{\bf Z}} \newcommand{\IR}{{\bf R}}
 \newcommand{\IT}{\bf T}
 
\newcommand{\IS}{\bf S} \newcommand{\IRP}{\bf RP}

\newcommand{\IZZ}{{\bf Z}_2}
\newcommand{\OMtp}{$OM2^+~$}
\newcommand{\OMtm}{$OM2^-~$}
\newcommand{\SLtz}{$SL(2,{\bf Z})~$}
\newcommand{\Ltb}{$L_{IIB}~$}
\newcommand{\Ttm}{${\bf T}^2_M~$}
\newcommand{\pq}{$(p,q)~$}
\newcommand{\wtilde}{\widetilde}
\newcommand{\Othm}{$O3^-~$}
\newcommand{\Othp}{$O3^+~$}
\newcommand{\Othmt}{$\widetilde{O3^-}~$}
\newcommand{\Othpt}{$\widetilde{O3^+}~$}

\newcommand{\wwwspires}{http://www.slac.stanford.edu/spires/find/hep/www}

\preprint{MIT-CTP-2946\\ TAUP-2609-2000\\ NSF-ITP-00-07\\ {\tt
hep-th/0003025}}

\title{On Orientifolds, Discrete Torsion, Branes and  M Theory}

\author{Amihay Hanany
\\
Center for Theoretical Physics,
\\ Massachusetts Institute of Technology\\ Cambridge MA 02139\\
\email{hanany@mit.edu}
}
\author{Barak Kol
\\
School of Physics and Astronomy\\ Tel Aviv University\\ Ramat Aviv 69978,
Israel\\
\email{barak1k@post.tau.ac.il}
}

\abstract{ We find some lifts to M theory of orientifold and
orbifold planes including the O1, O3 and O5 planes of Type
IIB and their transformations under \SLtz. The possible
discrete torsion variants (or K theory classes) are
explored, and are interpreted as arising from brane
intersections with planes. We find new variants of the O0
and of an orbifold line (OF1) and determine their tensions
in some cases. A systematic review of orientifolds, M
orientifolds, and known M lifts, with some new
clarifications is included together with a discussion of
the role of T duality.}
\keywords{M-theory, p-branes}
\begin{document}

\section{Introduction}

Orientifold planes are objects in string theory which are
defined perturbatively by gauging a discrete symmetry
which involves reversing the sign of coordinates
transverse to the plane while changing the orientation of
the string \cite{Sagnotti:1987tw,Horava:1989vt,GP}. Before
we delve into the details of their construction and their
dynamics let us mention some motivation and interest in
their study.

Orientifold planes are of interest in many aspects of
string theory. They turn out to be useful in studying
disconnected components in moduli space of various string
compactifications. They are crucial in the brane
construction of gauge theories with $Sp/SO$ gauge
groups.\footnote{See \cite{Clifford} for a recent
interesting application.} In addition they give a simple
realization of matter in symmetric and antisymmetric
second rank representations in $SU(N)$ gauge theories.

As will be discussed in detail in the following sections,
orientifold planes turn out to be characterized by
discrete fluxes. Some of these fluxes originate from NS
fields (See \cite{Zurab} for a recent discussion) and some
other fluxes come from RR fields. While the first (NS) are
visible from the usual perturbative description of
orientifolds, the latter (RR) cannot be treated in a
perturbative formulation of the Type II string since, as
of now, there is no formalism which includes RR
backgrounds in string theory. For this reason we need to
rely on other methods in order to study such objects, and
once such a formalism is available, these questions should
be revisited. The best analysis which is available at the
moment will include essentially two tools
- various dualities, and discrete fluxes represented by
intersections of branes with the orientifold plane.

Here we study two subjects related to orientifolds - their
lift to M theory (See a recent related discussion in
\cite{Angel}), and the different variants they appear in.
We begin with a review of the relevant features of
orientifolds in section \ref{review}. We describe the 4
types of $Op$ planes which are known to exist for $p \le
5$. We explain how these 4 types are classified by a pair
of $\IZZ$ parameters, one perturbative coming from the
$B_{NS}$ form and the other from one of the RR forms.
These parameters are ``discrete torsion'' parameters which
arise whenever the field strength of a p-form potential
has a corresponding non trivial discrete cohomology in
space time \cite{Vafa:1986wx,Vafa:1995rv} (see also recent
related work
\cite{Dijkgraaf:1999za,Gomis:2000ej,Klein:2000tf,Angelantonj:1999xf}).
Equivalently we show that these variants arise as brane
configurations in the spirit of \cite{HananyWitten} (see
recent related work \cite{Bergman:1999na}) from
intersections of an O plane with a brane which divides it
into two parts. We search for all the possible discrete
torsion variants, and we find that for $p \le 1$ there are
some additional ones. Later in that section, in
\ref{duality}, we explain how T duality relates a wrapped
orientifold with a pair of lower dimensional orientifolds
and the mapping between the various discrete torsion
variants.

In section \ref{revM} we systematically review the $\IZZ$
orbifolds of M theory. There are only a few of these and
upon compactification they give rise to the various $\IZZ$
orientifolds and orbifolds of string theory. We review the
M lifts of the O0, O2, O4
\cite{DasguptaMukhi,HananyKolRajaraman,
Sethi:1998zk,BerkoozKapustin,Witten_OM, Hori,Gimon:1998be}
in terms of 11d O planes - the OM1, OM2 and OM5. In the
case of the $O2$ we give a new M theory explanation for
the difference between the two variants $O2^+$ and
$\wtilde{O2}^+$.

In section \ref{Mlift} we describe the M lift of the O1,
O3 and O5 in terms of the same OM planes. In all cases we
pay special attention to the \SLtz symmetry
\cite{SchwarzSL2Z} (see also related work \cite{Kol5d}).
For the O3 we show how the M lift realizes the \SLtz
geometrically, as usual. For the O1 and O5 the S
transformation of \SLtz  transforms the orientifold planes
into orbifolds, which originate in the same OM planes.

In section \ref{orb} we discuss in detail the
interrelations between $Op$ planes and orbifolds for
$p=1,5$. We explore the possible discrete torsion variants
and discuss new ones for the orbifold and orientifold
lines and for the orientifold point. We compare the
discrete torsion classification with a recent claim on a K
theory classification \cite{MooreWitten} (see
\cite{Minasian:1997mm,Witten:1998cd,Horava:1999jy,
Bergman:1999ta,Gukov:1999yn,Hori:1999me}
for earlier and related work) and find groups of the same
order but with a different group structure. We are able to
find the tension of some of the variants, and it remains
to do so for the other cases.

Finally we discuss some miscellaneous applications in
section \ref{misc}. We discuss the dyon spectrum for 4d
${\cal N}=4$ with $SO/Sp$ gauge groups, where the $Sp$
theory has two versions. Then the discussion on the
monopole spectrum of such a theory is generalized to a
field theory in an arbitrary dimension. Finally, we
discuss the problems associated with a \pq web of
orientifolds.

Let us mention some open questions
\begin{itemize}
\item We show how an intersection of an orientifold, or an OM plane, with a
brane causes a tension jump, and so a {\it fractional} charge is
deposited on the brane. This jump should be understood and
probably required by the worldvolume theory on the brane. In
particular consistency with Dirac quantization is required.
\item We find the tension of some of the discrete torsion variants of the
 orbifold lines, and it would be interesting to know all of them.
\end{itemize}

Preliminary results of this
\href{http://online.itp.ucsb.edu/online/susy99/hanany/}{work} were
presented in Santa Barbara and can be found in the following link:
\href{http://online.itp.ucsb.edu/online/susy99/hanany/}
{http://online.itp.ucsb.edu/online/susy99/hanany/}.

\vspace{2cm}
\section{\protect\bigskip Introduction to Orientifolds}
\label{review}

An orientifold plane in $p+1$ dimensions is defined as
Type II string theory on $R^{p,1}\times
R^{9-p}/(I\cdot\Omega\cdot J)$ where $I$ is the inversion
of all coordinates in the transverse space $R^{9-p}$,
$\Omega$ is the orientation reversal on the world sheet of
the fundamental string and $J$ is the identity operator
for $p=0,1  ~{\rm mod}~4$ and $(-1 )^{F_L}$, the left
moving spacetime fermion number operator, for $p=2,3 ~{\rm
mod}~4$ \cite{DabholkarPark}. We will denote an
orientifold p-plane as an ``$Op$ plane''.

By definition the orientifold acts on $B_{NS}$ (with
components parallel to the $Op$ plane) with a sign
reversal. The action on the other NS fields is trivial, so
let us specify the action on all of the RR forms. The $O9$
which projects Type IIB to Type I keeps invariant the
left-right symmetric part of the spectrum, which in the RR
sector is the 2-form $B_{RR}$. By T dualizing we arrive at
the following rule for the action of an $Op$ plane on the
RR forms $C_{p'}$
\bea C_{p'} \to C_{p'} ~~p'=p+1~{\rm mod}~4  \nonumber \\
 C_{p'} \to - C_{p'} ~~p'=p+3~{\rm mod}~4 .
 \eea
Note that the above sign comes in addition to a component
dependent sign which comes from the tensor transformation
rules. Forms which get reversed by the action of the group
are termed ``twisted'', and this notation should not be
confused with forms from a twisted sector.

The transverse space to an $Op$ plane is an $\IRP^{8-p}$.
This space has some (discrete) torsion cohomologies which
are summarized in appendix \ref{cohomology}. Whenever a
space has discrete torsion, forms of an appropriate rank
are topologically classified by it. Namely
\be
[G_{p+1}] \in H^{p+1}(X,\IZ) \ee where $H^{p+1}(X,Z)$ is
the integral cohomology of the space $X$ and $G_{p+1}$ is
the RR field strength $p+1$-form. Twisted forms require
the use of twisted cohomologies, $\tilde{H}^{p+1}$. In the
case of orientifolds we will see that discrete torsion
allows for the orientifold to appear in several variants.

For a general orientifolds ($p \le 5$) there are at least
four types of orientifold planes which are distinguished
by two $\IZZ$ charges. One $\IZZ$ charge comes from the NS
sector and the other comes from the RR sector. For this
reason, the former charge is seen in perturbative string
theory while the latter is not. The perturbative
distinction was first discussed in modern language in
\cite{GP}. Later, the other $\IZZ$ charge was discovered
in \cite{Uranga,WittenAdSBaryons}\footnote{In
\cite{Witten6d} there is an earlier discussion of the four
variants of 5 planes in an M theory language which is
directly related to discrete torsions.} for the case $p=3$
essentially by using field theory intuition which is based
on Montonen Olive Duality or Type IIB self duality. The
case $p=4$ was studied in detail by
\cite{Hori,Gimon:1998be}.

For an $Op$ plane, the two $\IZZ$ charges are given by
$b$, which is the class $[dB_{NS}]=[H_{NS}] \in
\tilde{H}^3(\IRP^{8-p})$ and by $c$, which is the class of an
 RR form $[dC_{5-p}]=[G_{6-p}] \in \IZZ$
(the relevant cohomology, either twisted or untwisted, is
found to be $\IZZ$). In the next subsection we will give
these a physical interpretation via brane-orientifold
intersections. The above classes can be defined for $p \le
5$, while for $6
\le p \le 8$ one can still define $b$ through the
perturbative action on open strings.  In general the
existence of a discrete torsion implies the existence of
variants, though variants could possibly exist without it,
as in the case of the $b$ variant. The absence of a $c$
charge for $p=6$ suggests that, contrary to naive
expectations, one can not have an $\wtilde{O6}^-$ plane,
i.e. half a D6 brane stuck on an $O6^-$ plane. That turns
out to be the case as implied by the results of
\cite{Dijk}.

One can systematically check for any $p$ which additional
forms can get discrete torsions. For low $p$ some extra
variants exist (see section \ref{orb}): for $p
\le 1$ one can define an additional charge $c'$
by $[dC_{1-p}]=[G_{2-p}]
\in \IZZ$, and for $p=1$ one has $*H_{NS} \in
\tilde{H}^7(\IRP^7)=\IZZ$. In addition there are two other discrete
torsion whose variants we ignore - $H^0(\IRP^n)=\IZ$ leads
to a integral discrete variant for $p=2 ~{\rm mod}~ 4$
which we interpret as describing the possible massive
theories of type IIA (this interpretation is based on the
brane intersection picture, where this torsion is seen to
be due to an intersection with a D8 plane). The other type
which we ignore is $H^{8-p}=\IZ$ for odd $p$ and
$\tilde{H}^{8-p}=\IZ$ for even $p$, which simply means
that one can add to an $Op$ plane any integer number of
$Dp$ branes.

The orientifold planes will be denoted according to their
$\IZZ$ charges. A trivial $b$ charge will be denoted by a
$^-$ superscript and a non-trivial $b$ will be denoted by
$^+$. A non-trivial $c$ charge will be denoted by adding a
$\widetilde{}
\,$.
 The charges of these planes are $-2^{p-5}$
for the $Op^-$ plane, $+2^{p-5}$ for both the $Op^+$ and the
$\widetilde{Op^+}$ and ${\frac{1}{2}}-2^{p-5}$ for the
$\widetilde{Op^-}$ plane. The tensions of these objects are
measured in units of the Dp brane tension and are identical to
their charge. Note a change of notation for the various
orientifold planes from the papers \cite{HZ,HananyKolRajaraman}.
The orientifold planes are denoted there $Op^-, Op^+,
\widehat{Op},$ and $\widetilde{Op}$, respectively. The new
notation is based on the two $\IZZ$ charges and makes it more
simple to work with. $n$ physical $Dp$ branes stacked upon an $Op$
plane leads to a gauge group $SO(2n), Sp(n), SO(2n+1)$ and $Sp(n)$
for $Op^-, Op^+, \widetilde{Op^-}, \widetilde{Op^+}$,
respectively. The two $Sp(n)$ theories differ, in some cases, by a
theta angle and in their monopole spectrum. This will be discussed
in detail in sections \ref{spectrum} and \ref{BPS}.

\vspace{2cm}
\subsection{Brane Realization of Discrete Torsion}

\EPSFIGURE[h]{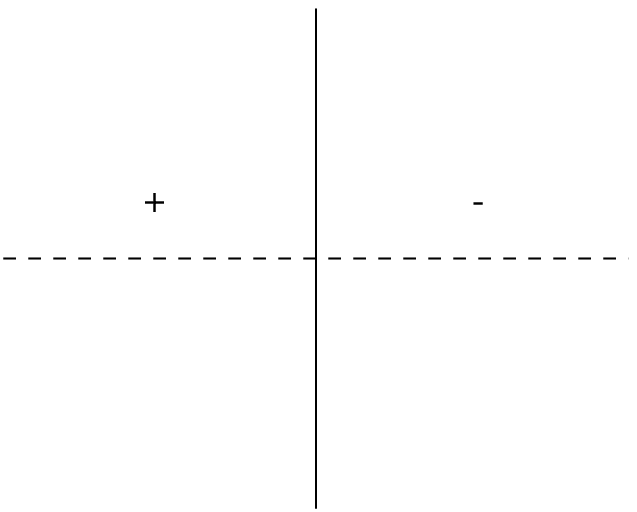}{A half NS brane and an Orientifold plane. The
solid line represents a half NS brane and the dashed line
represents an $Op$ plane. The type of the $Op$ plane
changes from $Op^-$ to $Op^+$ as it crosses the half NS
brane. \label{basic}}

In this section we will use a basic fact about orientifold planes
and NS five branes and will develop a set of relations between
intersecting branes and orientifold planes. The basic observation,
made in \cite{EJS}, can be described in figure \ref{basic}.
Consider a NS five brane which spans the 012345 coordinates and
intersecting an $Op$ plane which spans the $0,\ldots, p-1$, and 6
coordinates, $p\leq6$. It splits the orientifold plane into two
different parts, one to its left and one to its right and the
orientifold plane changes its type as in figure \ref{basic}.

The NS brane in this configuration has the special property that
it is reflected onto itself by the orientifold. In a compact
configuration such a brane would have half of the charge of a
brane which does not intersect the orientifold and so we call it a
${\frac{1}{2}}$ NS brane. Note that a 1/2 brane which extends
outside of the orientifold has the same charge density as a unit
brane, only it covers half of the volume due to the projection. On
the other hand, a half brane inside the orientifold has half the
charge density. A 1/2 brane cannot move in directions transverse
to the orientifold plane, because only integral branes can move
away off the orientifold. In the following brane configurations
one should bear in mind that all branes which intersect with an
orientifold plane are 1/2 branes.

\EPSFIGURE[ht]{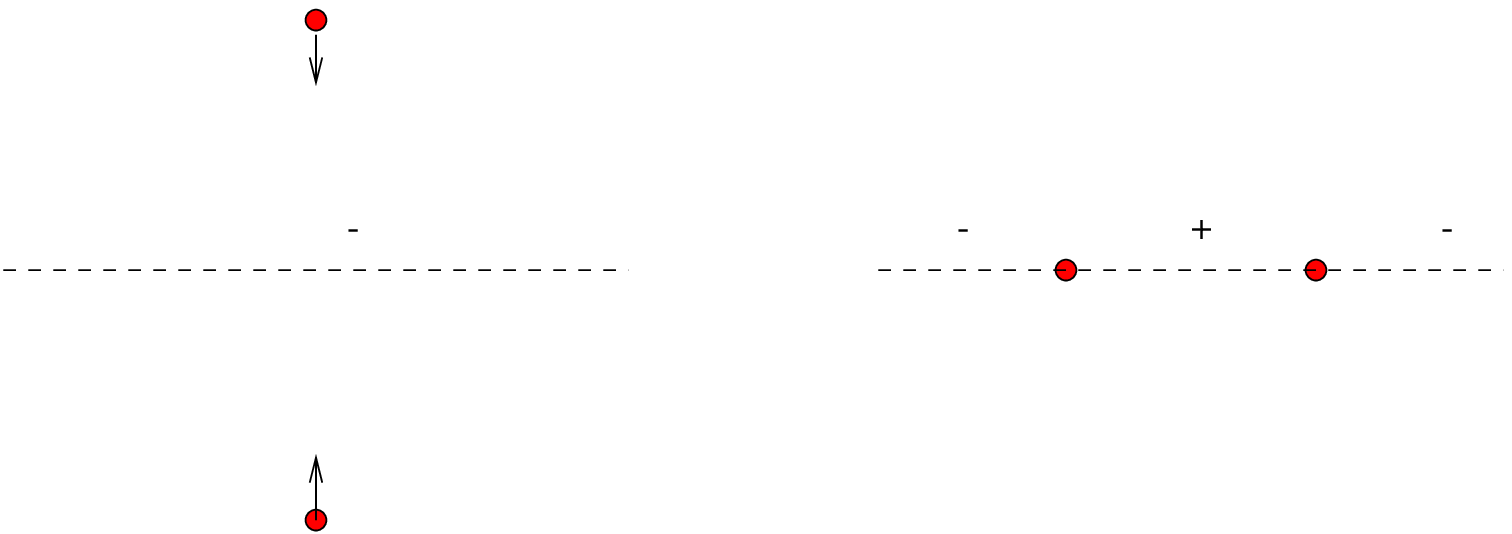}{A physical NS brane (and its image)
 near an orientifold plane. In the left
figure (a), the NS brane and its image are away from the
orientifold plane. As they move and meet along the $Op$
plane they can split into a pair of half NS branes and
form a new type of orientifold plane (b). The $Op$ types
are denoted by $+$ and $-$ signs.
\label{smooth}}

Let us emphasize this point in more detail. Consider a
configuration in which a physical NS brane is located far
away from the orientifold, as in figure \ref{smooth}a. Let
this brane move slowly towards the orientifold plane. It
moves together with its image under the orientifold
projection. As the two images meet on the orientifold
plane, they can split along it as in figure \ref{smooth}b.
At this point the rule in figure \ref{basic} can be used
and a new type of orientifold plane is generated in
between the two half NS branes.

We can look at it from a different point of view starting
from figure \ref{smooth}b and moving to figure
\ref{smooth}a. As observed in \cite{EJS}, the type of the
orientifold plane changes as one crosses the
${\frac{1}{2}}$ NS brane. This is determined by the $b$
charge of the orientifold which changes as one crosses the
${\frac{1}{2}}$ NS brane. On the other hand, if we add an
additional ${\frac{1}{2}}$ NS brane, the type of the
orientifold changes back to its original value, as in
figure \ref{smooth}b. In this case we can have a dynamical
process in which the two half NS branes combine together
along the 6 coordinate and leave the orientifold along the
789 directions as in figure \ref{smooth}a. This is the
inverse process to the one described in the last
paragraph.

One can relate such configurations to the value of the two form
flux $b$ by the following reasoning. The combined configuration of
a $Dp$ brane and half an NS brane is located at a point in 789
directions. There is a corresponding $\IRP^{2}$ which surrounds
the configuration. The field which couples magnetically to the NS
brane is the 2-form NS field. Consequently, the integral of the
two form over the $\IRP^{2}$ measures the number of 1/2 NS branes,
mod 2, which are located within the $\IRP^{2}$ \footnote{A
discussion in a similar spirit can be found in \cite{Tsabar}.}.
\begin{equation}
\exp\left(i\int_{\IRP^{2}}B_{NS}\right)=(-)^{\#\frac{1}{2} \rm{NS
~branes}}.
\end{equation}
This holonomy (``Wilson loop'') $\in H^p(X,U(1))$ of a
general p-form potential has the same discrete torsion as
the field strength class $\in H^{p+1}(X,Z)$ which was
mentioned before. This is seen via the exact sequences $0
 \to \IZ \to \IR \to U(1) \to 0$ and
\bea
\dots \to& H^{p}(X,\IR) \to& H^{p}(X,U(1)) \to \nonumber \\
\to H^{p+1}(X,\IZ) \to& H^{p+1}(X,\IR) \to&  \dots
\eea
where the cohomologies with real coefficients do not have
torsion parts.

We will assume that the configuration of figure
\ref{basic} exists and will apply some S and T dualities
to see what other results can be obtained from this basic
configuration. First let us apply S duality to this
configuration in the case $p=3$ (see also \cite{Mukhi3}).
For this we need to know what are the S transformations of
the various orientifold planes. This is easily done by
looking at their $\IZZ$ charges which are the NS 3-form
and the RR 3-form field strengths. Let us recall that
$O3^-$ has charge $(b,c)=(0,0)$ and the $\widetilde{O3^+}$
has charge (1,1). These orientifolds are self-dual under S
duality. $O3^+$ has (1,0) and transforms under S duality
to (0,1) which is the $\widetilde{O3^-}$.

Equipped with this data let us make an S duality on the
configuration of figure \ref{basic}. One gets that an
$O3^-$ transforms to an $\widetilde{O3^-}$ while crossing
a half D5 brane. By crossing another half NS brane the $b$
charge of the orientifold jumps and one gets
$\widetilde{O3^+}$. This is expected if one assumes that
the configuration of half D5 and half NS brane are self
dual under S duality and that the order of the D5 and NS
brane is not important. The summary is that when crossing
a half NS brane the $b$ charge changes while when crossing
a half D5 brane the $c$ charge of the orientifold changes.
$b$ measures the number of half NS branes enclosed by an
$\IRP^2$ surrounding the intersection of O3 plane and NS
brane while $c$ measures the number of half D5 branes
enclosed by an $\IRP^2$ surrounding the intersection of O3
plane and D5 brane.

Next we can perform T duality on these results. Consider
dualizing this system in a supersymmetric fashion to a
system of a NS brane along 012345 and a D6 brane along
0123789. The orientifold plane dualizes to an O4 plane
along 01236. The two $\IZZ$ charges are measured by $b$
and by $c$ which is a Wilson line of the RR one form of
Type IIA. T duality then implies that when an O4 plane
crosses a half NS brane its $b$ charge changes while when
crossing a half D6 brane its $c$ charge changes (A Wilson
line $c$ can be associated to our configuration by
repeating the arguments for the O3 plane). The system is
located at a point in the 45 directions. The object which
couples magnetically to the D6 brane is the RR 1-form of
Type IIA. Correspondingly $c$ measures the number of half
D6 branes, mod 2, trapped inside the $\IRP^1$.

Let us summarize the situation after further applying T
duality to other directions. An $Op$ plane along
012,\ldots,$p-1$ and 6, a half NS brane along 012345 and a
half D(p+2) brane along 012,\ldots,$p-1$, 789. {\it When
crossing a half NS brane the $Op$ plane changes its $b$
charge and when crossing a half D(p+2) brane the $Op$
plane changes its $c$ charge} which is measured by a
`Wilson loop integral' of the RR $(5-p)$ form potential.
This configuration exists for any $p$ which is 5 or less.
$b$ measures the number of half NS branes in an $\IRP^2$
enclosing the configuration while $c$ measures the number
of half $D(p+2)$ branes inside an $\IRP^{5-p}$ which
surrounds the configuration.

\subsection{T Duality}
\label{duality}

Let us apply T-duality in a direction along the orientifold plane.
An $O(p+1)$ plane wrapping a circle of radius $L$ turns after T
duality to a pair of $Op$ planes on a circle of radius $L'=1/L$.
Since we have (at least) 4 possible types for each $Op$ plane we
expect 16 possible types for the wrapped $O(p+1)$ plane.

Although this may sound surprising at first, we will
demonstrate the 16 possibilities by analyzing the possible
discrete torsions. The transverse space is $\IRP^{7-p}$
with cohomologies given in appendix \ref{cohomology}, and
we should find the 9 dimensional fields that can have
discrete torsions. Before wrapping the $O(p+1)$ plane the
fields with $\IZZ$ discrete torsions are $B_{NS}$ and
$C_{5-(p+1)=4-p}$. After compactifying on $L$ there are
two more such fields
- the reduction of the metric on $L$ (an untwisted 1
form), and the reduction of $C_{6-p}$ on $L$. To find the
relation with the T dual picture it is useful to list the
various discrete torsions in both pictures

\be
\begin{tabular}{|c||c|c|c|c|}
\hline
$O(p+1)$ & $B_{NS}$ & $C_{4-p}$ & $g_{\mu \nu}/L$ & $C_{6-p}/L$ \\
\hline
$O(p)$ & $B_{NS}$ & $C_{5-p}/L'$ & $B_{NS}/L'$ & $C_{5-p}$ \\
\hline
\end{tabular}
\label{Tduality}
 \ee
 where the notation $C_q/L$ means ``the form
$C_q$ reduced on the circle $L$''. Let us restate this
mapping in terms of brane intersections
 \be
\begin{tabular}{|c||c|c|c|c|}
\hline
$O(p+1)$ & NS5 & $D(p+3)$ & KK monopole & $\hat{D}(p+1)$ \\
\hline
$O(p)$ & NS5 & $\hat{D}(p+2)$ & $\widehat{NS}5$ & $D(p+2)$ \\
\hline
\end{tabular}
\label{TdualityBranes}
 \ee
 where the hat above a brane means that
it does not wrap the circle.

Let us discuss some examples. An NS5 intersecting the
$O(p+1)$ turns into an NS5 which intersects both of the
$Op$'s while an intersection with a $D(p+3)$ turns into a
$D(p+2)$ which intersects only one out of the pair. If we
want to get an RR discrete torsion on both $Op$'s then we
should intersect the $O(p+1)$ with a $D(p+1)$, one which
does not wrap the circle (and still has 4 mixed directions
relative to the O plane). If we take an $Op^-,~Op^+$ pair,
which has zero charge, then it transforms into an $O(p+1)$
intersecting a KK monopole on the circle $L'$ and must
have zero charge as well.

\section{Review of M-lifts of Orientifolds}
\label{revM} In this section we review known lifts of orientifold
planes to M theory. Such a lift requires lifting the
$\IZZ$ action to M theory. The objects we describe are
lifts of both orientifolds and orbifolds and accordingly
are denoted as $OMp$ planes. Since the worldsheet
formulation is lost in the lift, orientation reversal is
meaningless. Nevertheless we shall sometimes continue to
call them ``orientifolds''.

Lacking a fundamental definition of M theory, we are satisfied by
specifying the $\IZZ$ action on the 11d supergravity fields. An
$OMp$ plane includes a transverse spatial reflection, so we look
at M theory on ${\bf R}^{p,1} \times {\bf R}^{10-p}/ {\bf Z}_2$
where the first factor is the worldvolume of the $OMp$ plane, the
${\bf Z}_2$ in the second part is the reflection, and often it
will be more convenient to replace the last factor with ${\bf
T}^{10-p}/{\bf Z}_2$. The reflection determines the action on the
metric ($g \to g$). The action on the 3-form $C$ is determined by
requiring invariance of the topological term in the action $\int C
\wedge G \wedge G$, where $G=dC$, to be
\be
C \to (-)^p C.
\ee

Supersymmetry provides another constraint. When acting on
fermions, the inversion of $10-p$ coordinates squares to
the identity for $10-p=0,1 ~{\rm mod} ~4$ and to $(-)^F$
for $10-p=2,3 ~{\rm mod} ~4$ (this is a consequence of
$(\Gamma_1 \Gamma_2 ... \Gamma_n)^2=(-)^{n(n-1)/2}$). Thus
in order to have a supersymmetric orbifold and a $\IZZ$
action rather than a $\IZ_4$ action we require
\be
p=1,2 ~{\rm mod} ~4 ~.
\label{susyp}
\ee

We see that the $\IZZ$ objects intrinsic to M theory, are
the OM1, OM2, OM5, OM6 and OM9. It is no surprise that the
orientifolds intrinsic to M theory include a 2 plane and a
5 plane which we denote by OM2 and OM5. We shall review
the definition of these orientifolds and the way they were
used to find the M lift of the O4 and the O2. Then we look
for intersections of branes and orientifolds in M theory,
where it turns out that there is only one such
configuration: a $1/2$ M5 stuck on an OM2. After that we
review the lift of another Type IIA orientifold, the O0,
through an M theory object which we may call OM1.

We summarize here the results. The OM5 and OM2 carry charges given by
\be
\begin{tabular}{|c|c|c|}
\hline
$OM2^-$ & $OM2^+$ & OM5 \\
\hline
-1/16   & +3/16   & -1/2 \\
\hline
\end{tabular}
\ee where the $OM2^-, ~OM2^+$ are two discrete torsion variants of
the $OM2$ plane (which will be discussed later in
subsection \ref{OM5OM2}), and the charge is in units of
bulk OM2 and OM5 branes. The other planes support a
``twisted sector'' matter rather than charge (they are
neutral in 11d)
\bea \nonumber
\begin{tabular}{|c|c|c|}
\hline
$OM1$ & $OM6$ & OM9 \\
\hline
a chiral fermion & a 7d $SU(2)$ vector multiplet & a 10d $E(8)$ vector multiplet
\\
\hline
\end{tabular}
\eea

The OM6 and OM9 will not be discussed any further in this paper.
The OM6 is M theory on $\IT^4/\IZZ$ and the OM9 is a Ho\v{r}ava -
Witten plane \cite{HoravaWitten}.

\subsection{The OM5 plane}
The ``orientifold'' M on ${\bf T}^5/ {\bf Z}_2$ was studied in
\cite{DasguptaMukhi, Witten_OM}. The ``untwisted'' sector has a 6d
gravitational anomaly that can be canceled by 16 tensor
multiplets. Moreover, local anomaly cancellation would seem to
require adding a twisted sector of 1/2 a tensor multiplet at each
of the 32 fixed points. This problem is avoided by using a
different method of canceling the anomaly - assigning to each one
of the fixed points a charge
\be
Q(OM5)= -1/2, \ee in units of physical M5 branes. Note that while
we can usually put a half brane on top of an orientifold plane,
one cannot put half an M5 on an OM5 due to  M theory flux
quantization \cite{WittenFluxQuant}.

\subsubsection{M-lift of the O4}
\label{OM5/O4}

The OM5 can account for the O4 planes of Type IIA
\cite{Hori,Gimon:1998be} (see also \cite{Ahn:ads7}). When
lifting the $\IZZ$ action from Type IIA to M theory it is
required to specify the action on the circle $R_{11}$.
Topologically there are three possible actions: the
identity, a reversal and a shift through half a circle.
The identity is interpreted as an OM5 wrapped on $R_{11}$.
Reversal is not allowed since it would accumulate to an
inversion in 6 directions which would break supersymmetry
(\ref{susyp}). A shift is the same as a non-trivial
discrete torsion for the Type IIA 1-form $A$. It is
exactly the discrete torsion which is present in the
$\widetilde{O4}^{\pm}$ planes. So ``untilded'' ($c=0$)
orientifolds M-lift to $\IR ^5/
\IZZ$ while the ``tilded'' ones ($c=1$) M-lift to
$(\IR^5 \times \IS ^1)/ \IZZ$ with the $\IZZ$ acting on
the circle by a shift.

Let us look more at the shift orbifold $(\IR^5 \times
\IS^1)/\IZZ$. It has two kinds of minimal 1-cycles, one of
them wraps $\IS^1$ (``the circle'') and the other is a
straight line between points identified by the $\IZZ$
action (``antipodal line''). For points away from the
origin of $\IR^5$ the circle is smaller, but as we
approach the origin the antipodal line becomes smaller,
and at the origin itself it is half the size of the
circle.

We summarize the different O4 planes and their M theory origin
\begin{enumerate}
\item $O4^-$. Charge $Q=-1/2$. This is simply an OM5 wrapping $R_{11}$.
\item $\widetilde{O4^-}$. Charge $Q=0$. This is the smooth shift orbifold,
and as such indeed does not carry charge. (Recall that a bound
state of an OM5 and a half M5 is not allowed).
\item $O4^+$. $Q=+1/2$. This is an OM5 with a full stuck M5. (It is stuck
by imposing as Wilson loop an element of $O(2)$ which is not
connected to the identity \cite{Gimon:1998be}).
\item $\widetilde{O4^+}$. Charge $Q=+1/2$. This is the shift orbifold with
a stuck M5 at the origin on the circle of half radius.

The two types of $\widetilde{O4}$ planes were conjectured to be
related to the two elements of $\pi_4(Sp(n))= \IZZ$.
\end{enumerate}

\subsection{The OM2 plane}
The OM2 orientifold was studied in \cite{Sethi:1998zk}
(see also \cite{Ahn:ads4}). The charge can be found by
considering the interaction $-\int C \wedge I_8(R)$ for M
theory on $X=\IT^8 / \IZZ$. The effective charge is
$-\int_X I_8(R)=-\chi/24$. Although this space is
singular, we can define its ``resolved cohomologies'' by
adding to the invariant (untwisted) cohomologies an extra
256 cohomologies (in $H^{2,2}$) from the RR twisted
sector. This totals the Euler characteristic to 384 and so
the total charge to
-16. Dividing by 256, the number of fixed points, we find
$Q=-1/16$ in units of M2 charge.

This orientifold allows a variant due to a discrete flux. The
transverse space is ${\bf RP} ^7$ and the only field strength form in
M theory is the 4-form $G$, so we are interested in the cohomology
 $H^4({\bf RP} ^7,\IZ)=\IZZ$ (Appendix \ref{cohomology}). The discrete torsion
  adds a charge of a $+1/4= -{1 \over 2}\int_{{\bf RP}
  ^7} {C \over 2\pi} \wedge {G \over 2\pi}$ in units of M2.
 We denote the OM2 with trivial discrete torsion by $OM2^-$ and the
 one with non-trivial torsion by $OM2^+$. Their charges are summarized by
\bea
Q(OM2^-)=-1/16 \nonumber \\
 Q(OM2^+)=Q(OM2^-) +1/4=+3/16.
\eea

\subsubsection{The brane - orientifold intersection in M theory}
\label{OM5OM2}

The \OMtp can be realized by branes.
 To do that we put a half M5 brane on an
\OMtm plane. This is an analogue (the only one) of the Type II
configurations in section (\ref{review}).

Note that like the case of the OM5, one cannot attach a half M2
brane on top of an OM2 due to M theory flux quantization
 \cite{WittenFluxQuant}.

Once we compactify M theory we get (at least) two more
possibilities for brane intersections. One possible configuration
comes from lifting OF1 and D6 with charge jump of 1/16. It lifts
to a wound OM2 intersecting a KK6 - a Kaluza Klein monopole
(section \ref{o1}). Another configuration is the lift of an OF1
intersecting a D2, that is a wound OM2 intersecting with a
transverse M2 with a tension jump of $+1$ (sections \ref{o1-M},
\ref{o1}). An inspection of this example shows that the jump of
one unit of F1 charge is actually represented by a physical wound
M2 brane which is stretched, with its mirror, like a T shaped
brane (this is the M-lift of a double F1 ending on a D2
\cite{ASY}).

\subsubsection{M-lift of the O2}
\label{OM2}

Once one identifies the M theory objects, the \OMtm and
the \OMtp,  one can go on and find the M-lift of the O2
planes in Type IIA \cite{Sethi:1998zk,BerkoozKapustin}.
Actually, originally it must have been easier to go in the
opposite direction and determine the charges of the OM
planes from the O2 planes. Like the case of M-lifting the
O4, we need to lift the $\IZZ$ action to $R_{11}$. This
time it must be a reversal - it cannot be the identity
because of the susy constraint on $p$ (\ref{susyp}). So in
M theory there are actually two fixed planes located at
the two fixed points on the circle.

The different O2 planes and their M theory origin are described by
\begin{enumerate}
\item $O2^-$. Charge $Q=-1/8$. This is a pair of \OMtm
  planes: $-1/8=2 \times -1/16$.
\item $\widetilde{O2^-}$. Charge $Q=-1/8 ~+1/2=+3/8$. This is a pair
  of \OMtp planes: $-3/8=2 \times -3/16$.
\item $O2^+, ~\widetilde{O2^+}$. Both have $Q=+1/8$. This is a
  composite pair of an \OMtm with an \OMtp, $1/8=-1/16 ~+3/16$. The two possible
  O2's correspond to the possible ordering of the
  OM2's. This can be seen by intersecting an $O2^+$ with a D4
  brane. After lifting to M theory and using the intersection rule
  explained in section (\ref{review}), one finds that the $\widetilde{O2^+}$ has the
  reversed order of OM2's.
\end{enumerate}

\subsection{The OM1 line}
\label{OM1}

Let us consider the M-lift of the orientifold point, the O0, of
Type IIA. We take the action on $R_{11}$ to be the identity (a
reversal is not allowed by equation \ref{susyp}),
 so we consider
M theory on $R_{11} \times \IS ^1 \times \IR ^9/\IZZ$. A
computation of 2d Gravitational anomalies for M theory on
$\IT^9/\IZZ$ suggests that there is a chiral fermion on
every fixed line as an ``M theory twisted sector''
\cite{DasguptaMukhi,rey}. An independent evidence for the
existence of a chiral fermion on the fixed line comes from
the computation of the Witten index of $Sp(N), ~SO(N)$
matrix quantum mechanics
\cite{KacSmilga,HananyKolRajaraman}. We may call this line
an OM1 orientifold.

It is interesting to get the ``twisted sector'' which is
described above truly from a twisted sector of string
theory (see also \cite{DasguptaMukhi,DasguptaMukhi2}). In
order to get a 2d model one needs to compactify on an 8
manifold, and Type IIB on $\IT^8/
\IZZ$ has the right action on the fields (this is the
orbifold which we will call $OP1_B$ in section \ref{o1}).
Type IIA divided by $I_8 \cdot (-)^{F_L}$ would also do,
but we will stay with the more geometric example. The
$OP1_B$ has a twisted sector from the 4 form wrapping the
256 resolved $H^{2,2}$ cohomologies (\cite{orbifolds}
describes the computation of twisted sectors in general).
These scalars are chiral as they inherit their self
duality from the 4 form. Thus we get one chiral scalar for
each of the 256 fixed lines. Comparing with the 512 chiral
fermions of M theory on $\IT^9 / \IZZ$ we see that they
could match by bosonization provided the periodicity of
the scalars is at the free fermionic value, as it should.
This actually gives a nice realization of 2 dimensional
bosonization as implied by a lift to M theory.

As we consider here the OM1 on the $R_{11}$ circle there
are two possible boundary conditions for the fermion. The
Neveu-Schwarz boundary conditions correspond to an O0 with
trivial $RR$ discrete torsion ($c_{RR}=0$), and Ramond
corresponds to $c_{RR}=1$. This is verified by a
computation of the Casimir energy in the two cases, which
matches the O0 mass
\be
M(O0)= \pm 1/32
\ee
where the units are of momentum quanta along the circle.

Let us  summarize the different O0 planes and their M theory
origin
\begin{enumerate}
\item $O0^-$. Charge $Q=-1/32$. This is an OM1 with NS boundary
  conditions, and with integral momentum.
\item $\widetilde{O0^-}$. Charge $Q=-1/32 ~+1/2$. This OM1 has NS
  boundary conditions, but carries half-integer momentum.
\item $O0^+, ~\widetilde{O0^+}$. Both have $Q=+1/32$. These OM1's are
  in the Ramond sector, and there are two of them due to the zero
  mode which generates a degenerate ground state.
\end{enumerate}

\section{M lifts of Type IIB Orientifolds}
\label{Mlift}
In this section we describe (new) M-lifts of various
orientifold planes in Type IIB, while paying special
attention to the transformation properties under \SLtz.

We start by M-lifting the O3 plane, where we get a nice geometric/
microscopic realization of the \SLtz symmetry. Then we discuss the
O1 and O5 planes. Their S duals, which we call OF1 and ON5, are
constructed, and will be further discussed in section \ref{orb}.
The ON5 was already discussed in some works (see \cite{HZ} for a
recent review on this plane), and the OF1 was discussed in
\cite{Sen:1996na}.

The method is to recall the M theory origin of a Dp brane in terms
of M branes and then to find the analogous construction of an Op
plane in terms of the OM planes which were reviewed in the
previous section. The basic correspondence is between M theory on
$\IT^2_M$ and IIB on a circle of radius $L_{IIB}$, so we should
always distinguish two cases according to whether the Dp brane
wraps \Ltb or not. At weak coupling the \Ttm has a short side and
a long side, such that their ratio is the string coupling (when
the RR axion vanishes).

\subsection{O3}
Let us recall the M-lift of the D3 brane. A D3 which does
not wrap \Ltb is an M5 wrapping the torus, while a D3
which wraps \Ltb is the M2.

The four kinds of O3 planes have the following charges (in
D3 units)
\bea
 Q(O3^-) = -1/4, \nonumber \\
 Q(\widetilde{O3^-}) ~=Q(O3^+) ~= Q(\widetilde{O3^+}) = ~+1/4.
\eea

We start by lifting an O3 which wraps \Ltb, for otherwise
the circle
\Ltb is inverted as well and we get two O3 planes at the two fixed
points. Since the M-lift of a D3 brane which wraps
\Ltb is an M2 which does not wrap \Ttm, we should take an
OM2 which does not wrap \Ttm. Because of the compactness
of
\Ttm we are considering actually {\it four OM2 planes}.
The simplest possibility is to take four
\OMtm planes. One checks that the total charge $4 \times
(-1/16)= ~-1/4$ fits the $O3^-$ plane as expected.

\EPSFIGURE[ht] {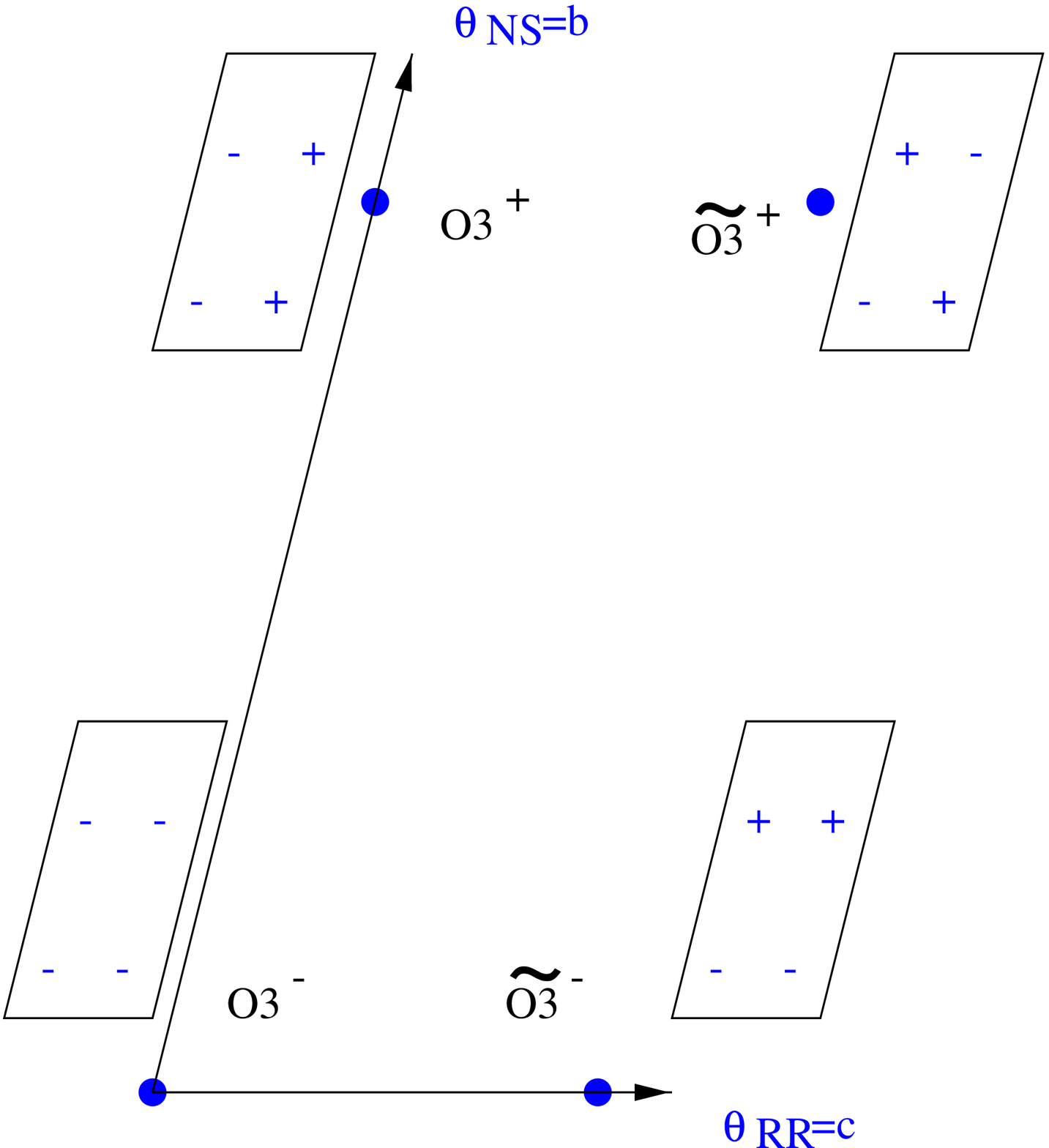,width=10cm}{The four types of $O3$ orientifolds
lifted to M theory. The $\pm$ signs stand for $OM2^\pm$
planes inside the M theory torus. \label{o3fig}}

To get the other O3 planes
\footnote{A.H. would like to thank Jacques Distler for discussions
on related points.}
 we use the brane intersection picture
of section \ref{review}. For example, to get the M-lift of
the $O3^+$ we should intersect the $O3^-$ with an NS5
brane. The lift of the NS5 is an M5 wrapping the long
side. Using the basic intersection in M theory (section
\ref{OM5OM2}) we see that we end up with two \OMtp planes
along the long side, and two other \OMtm planes, as in
figure \ref{o3fig}. The charges fit since $2 \times
(-1/16) + 2 \times (+3/16)=+1/4$. A similar argument works
for the $\widetilde{O3^-}$ and the $\widetilde{O3^+}$
planes by replacing the NS5 branes with a D5 brane or a
(1,1) brane respectively. Figure (\ref{o3fig}) summarizes
the various configurations.

Recall that the \SLtz properties of O3 planes can be described in
a diagram such as figure \ref{o3fig} \cite{Uranga,
WittenAdSBaryons}, where \SLtz acts on the torus in the diagram
according the its natural action on $(\IZZ)^2$. This action is
clearly visible from our M-lift into \Ttm.

There is an alternative way of finding the M lift of an O3
which uses T duality (section \ref{duality}). Under T
duality an O3 that wraps \Ltb turns into a pair of O2
planes, each one of which can be lifted to a pair of OM2
planes, as in section \ref{OM2}, giving 4 OM2 planes as
above. In this way one can recover the different lifts for
the different variants.

One may wonder about other choices for the signs of the
four OM2 planes. The ones with an odd number of signs
cannot be constructed by intersecting the $O3^-$ with a
Type IIB 5-brane as the 5-brane must intersect exactly two
OM2 planes (because the M5 is oriented). Nevertheless,
such configurations can be constructed making use of the
large but compact circle \Ltb, by intersecting the wound
O3 (wound on \Ltb) with a KK monopole (on \Ltb) as in
(\ref{TdualityBranes}). One gets a configuration with
three $OM2^-$ and one $OM2^+$ and total charge zero.
Intersecting now with an NS5 would give the other
possibility - one $OM2^-$ and three $OM2^+$. The case of 4
\OMtp planes is probably equivalent to four \OMtm planes
with an additional M2 brane.

Now we turn to an O3 which does not wrap \Ltb. It is
actually a pair of O3 planes, and we may T- Dualize them
into an O4 plane as in section (\ref{duality}). The M-lift
of the latter was described already in section
\ref{OM5/O4} in terms of OM5 planes.

\subsection{O5 - ON5}
Let us discuss O5 planes which wrap \Ltb, both because we
are less interested in a pair of O5's which we would have
had if the compact direction were inverted, and since we
are interested in configurations which lift to M5 branes
rather than KK monopoles.
\footnote{An alternative M lift
of the O5, the $ON5_B$ and their variants was given in
\cite{Witten6d}, using OM6 planes and more elaborate
quotients rather than OM5 planes.}
 For 5 branes, A \pq 5 brane which wraps \Ltb lifts to an
M5 wrapping a \pq cycle of
\Ttm. By analogy, we attempt to lift the O5 to an OM5
wrapping the short side of \Ttm. Such an OM5 plane is
actually a pair of OM5's because of the transverse compact
coordinate on \Ttm. We check that the charges match: $2
\times Q(OM5)= 2 \times (-1/2)= -1=Q(O5)$.

So far we have discussed the $O5^-$. We would like to
construct other discrete torsion variants of the O5, ones
which are independent of the compactification on
\Ltb, namely, those which are not related to forms that were
reduced on \Ltb. This can be done by turning on discrete
torsions for M theory on the torus.

Performing \SLtz we can get a family of \pq O5 planes,
such that the charge of a \pq $O5^-$ plane is $-1$ in
units of a \pq 5-brane. In particular we can consider a
$(0,1)$ O5 plane which we call $ON5_B$ because it is
charged under the same field that couples to the NS5 brane
of Type IIB (the charge is -1 in units of the NS5 charge).
The system of $ON5_B^-$ together with an NS5 brane can be
identified to be the $IIB/I_4 (-)^{F_L}$ orbifold and can
be called $ON5_B^0$ (section \ref{o5}; see \cite{HZ} for a
more detailed discussion). A set of N NS5 branes in the
vicinity of an $ON5_B$ results in 6d worldvolume gauge
theory, the same as a set of D5 branes near an O5 plane,
with the gauge group being one of $SO(2N),SO(2N+1),
~Sp(N)$ according to the type of the $ON5_B$.

Since NS5 branes exist both in Type IIA and in Type IIB, one might
expect the ON5 to exist in Type IIA as well. Indeed, the $IIA/I_4
(-)^{F_L}$ orbifold (section \ref{o5}), which we call an
$ON5_A^0$, is a system composed of an $ON5_A$ with an NS5 brane.
The $ON5_A^-$ has the property that when N NS5 branes coincide
with it, the worldvolume theory is a (2,0) CFT with a global
symmetry group $SO(2N)$, and this is the only possible variant.

\subsection{O1 - OF1}
\label{o1-M} The case of O1 is quite similar to the O5. Consider
an O1 which does not wrap \Ltb (so there is actually a pair of
O1's). As a D1 which does not wrap \Ltb M-lifts to a membrane
which wrap the long side of the torus, we should try an OM2 plane
wrapping the long side (so again there are actually two of them
because of the transverse short side). Let us check the charges:
$2 \times (-1/16)$ for the pair of O1's, indeed equals $2 \times
(-1/16)$ for the pair of OM2's.

For O1 planes which wrap \Ltb, we recall that  a D1
wrapping \Ltb is described by a unit of momentum along the
short side of \Ttm. So we try to wrap an OM1 along the
short side of \Ttm as its mass scales like units of
momentum (actually it is a pair of OM1's due to the
transverse long side). The charges work out for an $O1^-$
being made of a pair of $OM1^-$: $-1/16= 2 \times
(-1/32)$.

One can get other variants of the O1 by lifting brane intersection
to M theory. A new configuration happens for a pair of $O1^-$'s
which do not wrap \Ltb and are intersected by a D3 which does.
After the intersection we get a pair of $\wtilde{O3^-}$, and so
the tension jump is $2 \times +1/2= ~+1$. By lifting to M theory
we learn that the intersection of a wound OM2 with a transverse M2
gives a $+1$ tension jump.

Performing \SLtz we can get a family of \pq O1 planes,
such that the charge of a \pq $O1^-$ plane is the same as
for a \pq string. In particular we can consider a $(1,0)$
O1 plane which we call $OF1_B$ because it is charged under
the same field that couples to the fundamental string. We
will see that the $OF1_B$ can be identified with the
$IIB/I_8(-)^{F_L}$ orbifold (section \ref{o1}). Since
fundamental strings exist both in IIA and in IIB, one
might expect the OF1 to exist in Type IIA as well. Indeed,
this is the $IIA/(-)^{F_L}$ orbifold (section \ref{o1}),
and we call it an $OF1_A$. The perturbative orbifold
variants may be referred to as $OF1_B^0, ~OF1_A^0$.

\section{Orbifolds, Orientifolds and New Variants}
\label{orb}

Here we discuss the relations between orientifolds and
orbifolds and their variants. We start with lines, then 5
planes and then the O0. Throughout this section, when we
identify a perturbative orbifold with some plane which is
a dual of an orientifold, it should be borne in mind that
the identification holds only for one variant of the
plane, possibly with some extra matter, and all other
variants are produced by changing non-perturbative
discrete torsions.

\subsection{Orbifold lines}
\label{o1}
Orbifold lines together with the O1 form a family
connected by dualities. Table \ref{o1map} is our roadmap
for these connections. We will first explore this map and
then present some results on the tensions of discrete
torsion variants and a relation with K theory.

\TABLE{
\begin{tabular}{lccc}
          & \fbox{$IIB/I_8 ~\Omega ~~~~O1$} &   &\vspace{1mm}\\
          &  S $~~~\updownarrow$                    & & \\ \vspace{1mm}
\parbox{2cm}{(8,8) in 2d; OF (OM2)\\+variants}  & \fbox{$IIB/I_8 ~(-)^{F_L} ~~~~OF1_B$}&
$\longleftrightarrow ~~T \perp$ & \fbox{$IIA/I_8 ~~~~OF1_A$}
\vspace{1mm}\\
          & T $\parallel$ $~~~\updownarrow$   & & \vspace{1mm}\\
(16,0) in 2d;\\ OP (OM1)  & \fbox{$IIA/I_8 ~(-)^{F_L}
~~~~OP1_A$}& & \fbox{$IIB/I_8 ~~~~OP1_B$}\vspace{1mm}\\
\end{tabular}
\caption{Orbifold and orientifold lines. In this table the charges
and fluxes of the various O planes are not specified and may
change, depending on the particular case.} \label{o1map} }

Our starting point is the O1, that is, $IIB/I_8 \Omega$.
It carries D1 charge and the forms $B_{NS},C_0,C_4$ are
odd (twisted) under it. It has discrete torsion variants
due to $H_{NS},*H_{NS},G_5,G_1$ or in terms of brane
intersections due to the NS5, F1, D3 and a 7 brane. The 7
brane must allow a D1 charged object to end on it, and so
it should be a (0,1) 7 brane rather than a (1,0) D7.

We denote the S dual of the O1 by $OF1_B$ as S duality
replaces a D1 charge with an F1 charge. S duality replaces
$\Omega$ with $(-)^{F_L}$ and so this orbifold is $IIB/I_8
(-)^{F_L}$\footnote{To be more precise, one should note
that the $O1^-$, which is the O1 plane with no discrete
fluxes carries charge $-{1\over16}$, which must be
cancelled by adding non-perturbative discrete torsion and/
or extra matter, since the orbifold Type $IIB$ on
$\IT^8/\IZZ$ has 0 charge.}. The odd (twisted) forms under
the projection are $C_2,C_0,C_4,C_6$ and we get 16 (!)
discrete torsion variants from all four, or in terms of
branes from intersections with D1,D3,D5,D7. Since an F
string can end on any D brane, it is natural that each
intersection with a D brane is allowed and gives a new
variant.

Operating on the $OF1_B$ with a T duality in a direction
transverse to the fixed line gives an orbifold which we
denote by $OF1_A$, just as this operation acting on the F
string of IIB would give the F string of Type IIA. Such a
T duality is accompanied by an additional $(-)^{F_L}$, so
operating on $IIB/I_8 (-)^{F_L} \equiv OF1_B$ we get the
$IIA/I_8$ orbifold. This orbifold has 8 variants due to
intersections with D branes $D2,D4,D6$ or their respective
forms $G_2,G_4,G_6$. A D8 intersection is different
because the associated $G_0$ form has $H^0(\IRP^7)=\IZ$
cohomology rather than $\IZZ$ and is interpreted as a
change in the Type IIA cosmological constant. The D0 is
not in the list since it has no cohomology $H^8(\IRP^7)=0$
(but it may produce variants nevertheless).

Other orbifolds can be constructed now by compactifying an
OF1 line on a circle and performing parallel T duality.
This time one does not add an extra $(-)^{F_L}$. We get
$IIB/I_8$ and $IIA/I_8 (-)^{F_L}$. These orbifolds do not
have discrete torsion variants (when uncompactified). By T
duality they carry a momentum charge, so we denote them by
$OP1_A, OP1_B$.

The M lift of the O-lines can be found by looking at their
charges. After recalling the M lift of the F string we
conclude that the OF1 planes must be wrapping modes of the
OM2. The OP1 planes, on the other hand, are an unwrapped
OM1. Since the OM1 carries a chiral fermion (section
\ref{OM1}) after being compactified its Casimir energy
will give the required momentum charge.

Let us now find the \href{\hash ref-of1Tensions}{tensions} of
some of the discrete torsion variants of the OF1 lines
(table \ref{of1Tensions}). The tension of a bare OF1 is
$-1/16$ (in F string units) by S duality with the O1 . It
is consistent with the M description as an OM2 wrapping
the 11'th dimension, since the OM2 has tension $-1/16$ (in
M2 units).

\TABLE{
\begin{tabular}{|c|c|c|c|c|c|c|}
\hline
D1 & D2 & D3 & D4 & D5 & D6 & D7 \\
\hline
+2? & +1 & +1/2 & +1/4 & +1/8 &  +1/16? & +1/32? \\
\hline
\end{tabular}
\caption{Tension jumps of the OF1 after various intersections.
Question marks denote charges for which more consistency
checks are needed.} \label{of1Tensions} }
\begin{itemize}
\item To compute the tension of an $OF1_B$ after intersecting a D5, consider
performing S duality to an O1 which upon intersecting an
NS5 changes from $O1^-$ of tension $-1/16$ to an $O1^+$ of
tension $+1/16$ (so the jump is $+1/8$).
\item The intersection of an $OF1_A$ with a D4 can be M lifted to the basic
intersection of an OM2 with an M5 (section \ref{OM5OM2}),
and so the tension jump is $+1/4$.
\item The tension jump of an $OF1_B$ intersecting with a D3 is found again by S duality to
be $+1/2$.
\item The tension jump of an $OF1_A$ intersecting a D2 is
$+1$. This is a consequence of the M theory configuration
found in section \ref{o1-M}, where a wound OM2 intersects
a transverse M2. Since the tension jump is integral there
is nothing to prevent a whole F1 to separate from this
variant.
\item We seem to get a rule that after intersecting a Dp brane the jump is
$2^{2-p}$, which is consistent with T duality. For the D6
this rule has an independent check. A D6 intersection
corresponds after an M lift to a shift in $R_{11}$. So
this $OF1_A$ variant lifts to M theory on the smooth
manifold $(\IT^8 \times \IS^1)/\IZZ$ where the $\IZZ$ acts
by inversion on the first factor and by a shift on the
second. As a smooth manifold it carries zero tension, in
agreement with a $-1/16$ jump.
\item
The intersection of an $OF1_A$ with a D0 does not have a
discrete cohomology as $H^8(\IRP^7)=0$ so it is not clear
whether it gives a new variant. If there is a new variant
corresponding to intersection of the OF1 with D0 the jump
in its charge is 4. Conservation of the fundamental string
charge then implies that 4 physical fundamental strings
must enter the D0 together with the $OF1^-$ plane. This is
indeed consistent with the ``fork'' configuration of
\cite{Berlin} which gives some support for its existence.
\end{itemize}

One can discuss OF1 orbifolds with a discrete torsion from
several forms turned on at the same time. It would be
interesting to find their tensions.

The above variants have an interesting relation with
\underline{K theory}. It is simpler to consider $OF1_A=
~IIA/I_8$. Discrete torsion variants are classified by the
reduced cohomology $H^*(\IRP^7)=\IZZ \oplus \IZZ \oplus
\IZZ$. However, it was recently claimed that the correct
classifying group is the reduced K group
\cite{MooreWitten} $K(\IRP^7)=\IZ_8$ (``reduced'' simply
means in both cases that we do not write down a trivial
$\IZ$ factor). The K group is actually a ring which
differs from the standard $\IZ_8$ and is defined by the
following relations on its generator $x$
\bea
8x=0 \nonumber \\
 x^2 =-2x
\eea
Note that both the cohomology and the K ring have the same order (8) whereas
their structures differ. It would be interesting to elucidate the role of the
algebraic structure.

\subsection{Orbifold 5 planes}
\label{o5}
We can take a similar tour of orbifold 5 planes, with
table \ref{O5s} as our road map.

\TABLE{
\begin{tabular}{lccc}
          & \fbox{$IIB/I_4 ~\Omega ~~~~O5$} &   &\vspace{1mm}\\
          &  S $~~~\updownarrow$                    & & \\ \vspace{1mm}

\parbox{2cm}{(1,1) in 6d\\+variants}  & \fbox{$IIB/I_4 ~(-)^{F_L} ~~~~ON5_B$}&
 $\longleftrightarrow ~~T \perp$ & \fbox{$IIA/I_4 ~(OM6)$} \vspace{1mm}\\

          & T $\parallel$ $~~~\updownarrow$   & & \vspace{1mm}\\

(2,0) in 6d  & \fbox{$IIA/I_4 ~(-)^{F_L} ~~~~ON5_A ~(OM5)$}&  &
\fbox{$IIB/I_4$}\vspace{1mm}\\
\end{tabular}
\caption{Orbifold and orientifold 5 planes}
\label{O5s}
}

We start with the O5 orientifold. It has charge -1 in
units of D5, and it has a pair of discrete torsions due to
the forms $G_1,H_{NS}$, or in terms of branes due to
intersection with $D7, NS5$.

S duality creates the orbifold $IIB/I_4 (-)^{F_L}$, which
we denote by $ON5_B$ since it carries NS5 charge. In order
to cancel the charge one needs to add to the $ON_5$ plane
an NS5 and this is actually the perturbative orbifold. The
matter living on the NS5 of Type IIB is, of course, in
(1,1) 6d multiplets.  It has variants from the RR forms
$G_1,G_3$, allowing all the $SO,Sp$ gauge groups.

Performing T duality parallel to an $ON5_B$ gives Type
IIA/$I_4 (-)^{F_L}$, which we denote by $ON5_A$, because
it has NS5 charge as well. It carries matter in (2,0) 6d
multiplets. If N NS5 branes are added to the orbifold we
get a (2,0) theory with $SO(2N)$ group.

We can also perform T duality transverse to the orbifold
planes. We get $IIA/I_4$ and $IIB/I_4$. The first orbifold
has a variant due to $G_2$. These orbifolds are known to
carry non perturbative matter
- a (1,1) theory in the first case, and (2,0) in the
second.

Let us consider the M lifts of Type IIA orbifold 5-planes.
We can imagine two M theory orbifolds that would give us 5
planes in Type IIA, either an OM6 wound on $R_{11}$ or an
unwound OM5. By comparing the action on the fields we find
that the $ON5_A=IIA/I_4 ~(-)^{F_L}$ lifts to the unwound
OM5 while $IIA/I_4$ lifts to a wound OM6. The case of Type
IIB can be discussed as well, but it has more detail
because one needs to specify whether the 5 plane wraps
\Ltb or not.

\subsection{New variants of the O0 plane}
For the low dimensional orientifolds, the O0 and the O1, a
discrete torsion analysis predicts the existence of more
than two $\IZZ$ parameters. For the O1 we saw that there
are four $\IZZ$ parameters. In addition to the usual
$(b,c)=H_{NS},G_5$ there are also a pair of $\IZZ$'s from
$G_1$ and $*H_{NS}$.

Similarly, for the O0 we can analyze the possible discrete
torsions. In addition to the expected pair of $\IZZ$ parameters
$(b,c)=H_{NS},G_6$, there is a third $\IZZ$ from $G_2$. After an M
lift, this additional discrete torsion is nothing but a shift in
$R_{11}$, the possibility which was not considered section
\ref{OM1}. It would be interesting to find the mass of the 4
variants with $[G_2] \ne 0$.

\section{Miscellaneous applications}
\label{misc}
\subsection{The spectrum of 4d ${\cal N}=4$ with $SO, ~Sp~$ gauge group}
\label{spectrum}
We know that every orientifold 3-plane gives rise
to a 4 dimensional gauge theory on D3 probes parallel to it. Each
O3 plane gives rise to a theory with a different gauge group $G$,
according to
\begin{enumerate}
\item For $O3^- ~G=SO(2N)$.
\item For $\wtilde{O3^-} ~G=SO(2N+1)$.
\item For both $O3^+, ~\wtilde{O3^+}$ the gauge group
is $G=Sp(N)$.
\end{enumerate}

Since both \Othp and \Othpt have the same gauge group one may ask
how do the two theories differ. It is clear that the theory with
\Othpt is an \SLtz transform of the one with \Othp by the element
\be
T=\left[\begin{array}{cc}
 1 & 1 \\
 0 & 1
 \end{array}\right].
\ee

   We would like to show how this difference manifests itself in
one of the basic measurables of the theory, the spectrum of 1/2
BPS states.

By Definition we know that the W bosons lie in the root
lattice of the gauge group $G$. Knowing the lattice of the
W bosons and the
\SLtz transformation which acts naturally both on the
charges of the states and on the discrete charges of the
orientifolds allows us to characterize the 1/2 BPS
spectrum as follows
\be
\begin{tabular}{|r|c|c|c|c|}
\hline
                & \Othm & \Othmt & \Othp & \Othpt \\
        $(b,c)$= & (0,0) & (0,1)  & (1,0) & (1,1) \\
\hline \hline
\pq=(1,0) mod 2 & D     & B      & C     & C      \\
\pq=(0,1) mod 2 & D     & C      & B     & C      \\
\pq=(1,1) mod 2 & D     & C      & C     & B      \\ \hline
\end{tabular}
\ee Here $D$ denotes that the states lie in a $D=SO(2N)$ lattice,
$B$ is a $B=SO(2N+1)$ lattice and $C$ is a $C=Sp(N)$ lattice. We
get that the charge lattice is of type B iff $(p,q)=(c,b) ~\rm{mod
~2}$.

Note that the difference between \Othp and \Othpt is manifest in
their spectrum of monopoles and dyons - whereas the monopoles of
the \Othp theory lie in the $B$ lattice (the dual lattice), the
monopoles of the \Othpt theory lie in a $C$ lattice just like the
W bosons.

\subsection{A comment on allowed BPS stated and $\IZZ$ charges}
\label{BPS}

Consider a configuration with an $Op$ plane and a physical Dp
brane away from it. Our aim is to ask what are the allowed BPS
configurations which can stretch in between the Dp brane and its
image. For the simplest case, with $Op^-$ the gauge group is
$SO(2)$. It is known that a fundamental string stretched between
the Dp brane and its image does not lead to a BPS state but
rather, as Sen shows in detail in \cite{Sen}, to a non-BPS state.
This happens because the BPS ground state is projected out, so the
lowest state is the next massive level which is not  BPS.

We would like to extend this discussion to monopole
solutions and, when possible, to dyonic states. The first
example is for $p=3$. Using the results of the previous
subsection we find that the allowed BPS states are
described by the following table
\be
\begin{tabular}{|c|l|l|l|l|}
\hline (b,c) & (0,0) & (1,0) & (0,1) & (1,1) \\ \hline \hline
Fundamental String & no & yes & no & yes \\ D1 brane & no & no &
yes & yes \\ \hline
\end{tabular}
\ee

It is easy to see from this table that a brane is allowed
as a BPS state whenever the $\IZZ$ of the two form which
couples to it electrically is non trivial.

This leads to the following generalization for any $p$. Consider a
Dp brane and its image under a reflection by an $Op$ plane. Then
BPS states arise when either a fundamental string or a Dp-2 brane
is stretched between the brane and its image, according to the
following table:
\be
\begin{tabular}{|c|c|c|c|c|}
\hline (b,c) & (0,0) & (1,0) & (0,1) & (1,1) \\ \hline \hline
Fundamental String & no & yes & no & yes \\ Dp-2 brane & no & no &
yes & yes \\ \hline
\end{tabular}
\ee

Unlike the previous case, the simple rule that a brane is
allowed to be stretched in between the heavier brane and
its image whenever the corresponding $\IZZ$ flux of the
form which couples to it electrically is non-trivial does
not apply here. It is not clear how the picture
generalizes.

Similar statements hold for D3 branes stretching between
NS5 branes in the presence of the different types of
$ON5_B$ planes.

\subsection{Orientifold webs}
Since we identified \pq O1 lines and \pq O5 planes one may wonder
whether \pq webs of orientifolds are possible. We would like to show
that those may be possible in some special cases, but in general
they do not make sense.

Let us consider the basic junction at $\tau=i$, where
$\tau$ is the complex scalar of Type IIB. We have a
horizontal $(0,1)$ O1 meeting a vertical $(1,0)$ OF1 and a
$(1,1)$ O line which leaves the junction at 45 degrees.
Each orientifold plane requires a $\IZZ$ projection. If we
ignore the orientation reversal and consider only the
spatial inversion we see that these 3 reflections generate
a group of order 8, isomorphic to $D_4$, the dihedral
group of 4 elements. It is not clear how do the different
orientation reversals combine.

However, if there are two O lines at an irrational relative angle
$\alpha$ (measured in radians/$2\pi$) then a composition of both
reflections gives a rotation by $2\alpha$. Since we assumed
$\alpha$ irrational, then this element generates an infinite group
of identifications on the plane, generating a non-discrete image
set from a single point. Thus we cannot hope to have an ordinary
orbifold/orientifold, and this is the case for generic $\tau$.

\vspace{1cm}
\noindent {\large {\bf Acknowledgements}}
\vspace{.5cm}

We thank O. Bergman, E. Gimon, S. Gukov, E. Shustin, C. Vafa, E.
Witten and A. Zaffaroni for enjoyable discussions.

A.H. would like to thank the Institute for Theoretical
Physics at Santa Barbara and Tel-Aviv University for their
kind hospitality while various stages of this work were
completed. B.K. would like to thank J. Sonnenschein, S.
Yankielowicz and the rest of the group at Tel Aviv.

A. H. is partially supported by the National Science
Foundation under grant no. PHY94-07194, by the DOE under
grant no. DE-FC02-94ER40818, by an A. P. Sloan Foundation
Fellowship and by a DOE OJI award. The research of BK was
supported in part by the US-Israeli Binational Science
Foundation, the German--Israeli Foundation for Scientific
Research (GIF), and the Israel Science Foundation.

\appendix
\section{Appendix - Cohomology of $\IRP^n$}
\label{cohomology}
We distinguish between two kinds of cohomologies. The twisted
ones, denoted by $\tilde{H}$ classify ``twisted'' forms. These are
forms which reverse sign under the projection (such forms are not
related to a twisted sector). The ordinary ``untwisted''
cohomologies are denoted just by $H$ and are appropriate to
classify forms which do not change sign under the projection.

The integral cohomologies of $\IRP^n$ are
\bea
H^q =&& \left\{
\begin{array}{ll}
\IZZ & \mbox{ $q$ even but } q \neq 0\\
 \IZ & q=0, \mbox{ and for odd n } ~q=n \\
 0   & \mbox{otherwise}
\end{array} \right. \nonumber \\
\tilde{H}^q =&& \left\{
\begin{array}{ll}
\IZZ & \mbox{$q$ odd}\\
 \IZ & \mbox{for even n } ~q=n \\
  0 & \mbox{otherwise}
\end{array} \right.
\eea

These results can be easily deduced from the chain complexes
\be
\begin{array}{c}
\leftarrow
C^2 =\IZ
\stackrel{\times 2}{\longleftarrow}
C^1 =\IZ
\stackrel{0}{\longleftarrow}
C^0 =\IZ \leftarrow 0 \\
\stackrel{\times 2}{\longleftarrow}
C^{2m+1} =\IZ
\stackrel{0}{\longleftarrow}
C^{2m} =\IZ
\stackrel{\times 2}{\longleftarrow}
\dots \\
0 \leftarrow C^n =\IZ
\leftarrow
\dots
\end{array}
\ee
and
\be
\begin{array}{c}
\leftarrow
\tilde{C}^2 =\IZ
\stackrel{0}{\longleftarrow}
\tilde{C}^1 =\IZ
\stackrel{\times 2}{\longleftarrow}
\tilde{C}^0 =\IZ \leftarrow 0 \\
\stackrel{0}{\longleftarrow}
\tilde{C}^{2m+1} =\IZ
\stackrel{\times 2}{\longleftarrow}
\tilde{C}^{2m} =\IZ
\stackrel{0}{\longleftarrow}
\dots \\
0 \leftarrow \tilde{C}^n =\IZ
\leftarrow
\dots
\end{array}
\ee
where $C^q$ are the q-cochains and $\tilde{C}^q$ are the twisted q-cochains.

For completeness we list the integral homology groups as
well
\bea
H_q =&& \left\{
\begin{array}{ll}
\IZZ & q \mbox{ odd but }  q \neq n\\
 \IZ & q=0, \mbox{ and for odd n } ~q=n \\
 0   & \mbox{otherwise}
\end{array} \right. \nonumber \\
\tilde{H}_q =&& \left\{
\begin{array}{ll}
\IZZ & \mbox{$q$ even } q \neq n\\
 \IZ & \mbox{for even n } ~q=n \\
  0 & \mbox{otherwise}
\end{array} \right.
\eea

For odd $n$ $\IRP^n$ is orientable and Poincare duality
holds
\be
H_i=H^{n-i}, ~~~~~~ \tilde{H}_i=\tilde{H}^{n-i}
\ee
For even $n$, on the other hand
\be
H_i=\tilde{H}^{n-i}, ~~~~~  \tilde{H}_i=H^{n-i}
\ee

\bibliographystyle{JHEP}

\end{document}